\documentclass[12pt]{article} 
\newcommand{\nin}{\noindent} 
\newcommand{\eq}{\begin{equation}} 
\newcommand{\eqx}{\end{equation}} 
 
\begin{document} 

\pagestyle{empty} 
%
\begin{center} 
{\Large \bf Shake-up and shake-off excitations with associated electron
losses in X-ray studies of proteins }\\ 
\vspace{10mm} 

{\Large Petter Persson $^{1}$} , 
{\Large Sten Lunell $^{1}$}, 
{\Large Abraham~Sz\"{o}ke $^{2,\,3}$},\\ 
{\Large Beata~Ziaja $^{2,\,4,\,5}$} 
{\Large and Janos~Hajdu $^{2}$
\footnote[6]{e~-mail:petter.persson@kvac.uu.se,~sten.lunell@kvac.uu.se,~szoke1@llnl.gov,\\
ziaja@tsl.uu.se,~hajdu@xray.bmc.uu.se} }

\vspace{3mm} 
$^{1}$ \it Department of Quantum Chemistry, Uppsala University,\\ 
\it Box 518, S-751 20 Uppsala, Sweden\\ 

\vspace{3mm} 
$^{2}$ \it Department of Biochemistry, Biomedical Centre,\\ 
\it Box 576, Uppsala University, S-75123 Uppsala, Sweden\\ 

\vspace{3mm} 

$^{3}$ \it Lawrence Livermore National Laboratory, Livermore, 
CA 94551, USA\\ 
\vspace{3mm} 

$^{4}$ \it Department of Theoretical Physics, 
Institute of Nuclear Physics, 
\it Radzikowskiego 152, 31-342 Cracow, Poland\\ 
\vspace{3mm} 
$^{5}$ \it High Energy Physics, Uppsala University, 
P.O. Box 535, S-75121 Uppsala, Sweden 
\end{center} 
\vspace{5mm}
\nin
{\bf Corresponding author:}\\
Janos Hajdu, Department of Biochemistry, Biomedical Centre,\\
Box 576, Uppsala University, S-75123 Uppsala, Sweden\\
Tel:+4618 4714999, Fax:+4618 511755, E-mail:hajdu@xray.bmc.uu.se \\ \\
\nin
{\bf Manuscript information:}\\
Number of pages (double-sided version with tables and figure legend on
separate pages): 19 , Number of figures:1, Number of tables:2\\ \\
\nin
{\bf Submission information:}\\
Diskette of the manuscript included, manuscript written in LaTeX, ASCII
file
extracted; hard copy double-spaced\\
\newpage
\thispagestyle{empty}
\begin{abstract} 

Photoionisation of an atom by X-rays usually removes an inner-shell
electron from the atom, leaving behind a perturbed "hollow ion" whose
relaxation may take different routes. In light elements, emission of an
Auger electron is common. However, the energy and the total number of
electrons released from the atom may be modulated by shake-up and
shake-off effects. When the inner-shell electron leaves, the outer-shell
electrons may find themselves in a state that is not an eigen-state of the
atom in its surroundings. The resulting collective excitation is called
shake-up. If this process also involves the release of low energy
electrons from the outer shell, then the process is called shake-off. It
is not clear how significant shake-up and shake-off contributions are to
the overall ionisation of biological materials like proteins. In
particular, the interaction between the out-going electron and the
remaining system depends on the chemical environment of the atom, which
can be studied by quantum chemical methods. Here we present calculations
on model compounds to represent the most common chemical environments in
proteins. The results show that the shake-up and shake-off processes
affect about 20\% of all emissions from nitrogen, 30\% from carbon, 40\%
from oxygen, and 23\% from sulphur. Triple and higher ionisations are rare
for carbon, nitrogen and oxygen, but are frequent for sulphur. The
findings are relevant to the design of biological experiments at emerging
X-ray free-electron lasers. 
\end{abstract} 
\nin
\pagebreak 

\pagestyle{myheadings}
\markboth{Persson et al.}{Persson et al.} 

Computer simulations (Neutze {\em et~al.}, 2000) show that ultrashort 
and high-intensity X-ray pulses, as those expected from presently developed 
free-electron lasers (Winick, 1995; Wiik, 1997), may provide structural 
information from large protein molecules and assemblies before radiation 
damage destroys
them. Estimation of radiation damage as a function of X-ray photon energy,
pulse length, integrated pulse intensity and sample size was obtained in
the framework of a radiation damage model, where the effects of
atom-photon and ion-ion interaction were taken into account. Photons of
$1$ \AA $\,\,$ wavelength, corresponding to a photon energy of about 12
keV, interact with atoms mainly via the photoelectric effect (Dyson, 1973)
(for carbon the photoelectric cross-section is $\sim 10$ times higher than
the corresponding elastic cross-section at this wavelength (Hubbel 
{\em et~al.}, 1980)), and thus the photoelectric effect is the main source 
of radiation damage with X-rays. 

Photoionisation may proceed either through the ejection of an outer-shell
electron or through the ejection of an inner-shell electron (Dyson, 1973).
Outer-shell photo-events eject a single electron from the atom
with an energy equivalent to the energy of the incoming photon minus the
shell binding energy and the recoil energy. With photon energies of around
12 keV (or about 1 \AA$\,$  wavelength), outer-shell events are rare in carbon,
nitrogen, oxygen and sulphur, and represent less than 5\% of all
photoionisation events. 

A more frequent type of photoionisation with X-rays involves the ejection
of an inner-shell electron from the atom (around 95-97\% of
photoionisations remove a K-shell electron from carbon, nitrogen, oxygen
and sulphur). Typical K-hole life times are of the order $1-10$ fs
(Krause \& Oliver, 1979), and the hollow ion relaxes through an electron 
falling from a higher shell into the vacant hole. In light elements, the 
energy of this
electron is given to another electron which is then also ejected from the
atom through the Auger effect. Core ionisation, however, constitutes a
strong perturbation of the molecule, and may be accompanied by significant
electronic effects (Siegbahn {\em et~al.}, 1969). Firstly, the valence 
electrons often relax
significantly to compensate for the presence of the positive core
hole. The departing photoelectron can interact with these relaxing
electrons, and lose kinetic energy in the process, thus forcing the system
into an excited state (a process called {\em shake-up}). In some cases,
the excitation may result in the ejection of low energy outer-shell
electrons from the atom (a process called {\em shake-off}). The multiple
excitation lifetimes for light elements are comparable to core-hole
lifetimes. The interaction between the out-going electron and the
remaining system depends on the chemical environment of the atom, and a
description of these interactions requires quantum chemical
calculations. The following key processes need to be taken into account:

(i) PHOTOEMISSION FOLLOWED BY SINGLE AUGER DECAY.\\
Ejected K-shell
electrons with the highest possible kinetic energy correspond to a
situation where the remaining system is left in the ground state, i.e. the
difference between the incoming photon energy and the energy of the
ejected photoelectron equals the chemical binding energy and the recoil
energy. In light elements like
carbon, nitrogen, oxygen and sulphur, such clean photoemissions are
followed by Auger emission, and the atom becomes doubly ionised (Krause
\& Oliver, 1979). 
In heavier elements X-ray fluorescence dominates. 

(ii) SHAKE-RELATED PROCESSES. When photoionisation ejects an inner-shell
electron so fast that the outer-shell electrons have no time to relax, the
situation gets similar to beta decay, whereby the nuclear charge suddenly
increases by one unit. Quantum mechanics describes such a state as a
superposition of proper eigen-states, that include states where one or
more of the electrons may be unbound. Interactions between the departing
photoelectron and the electrons left behind may reduce the kinetic energy
of the photoelectron, and deposit energy into the system. The perturbation
is called shake-up if they refer to an excitation in the final system, or
shake-off if the result is the loss of one or more outer shell electrons
from the ion. The relative contributions from these two processes have
been found to be comparable in noble gases (Svensson {\em et~al.}, 1988;
Armen {\em et~al.}, 1985; Wark {\em et~al.}, 1991). 
The following shake-related phenomena need to be considered:

(a) Photoemission accompanied by shake-up excitation with Auger
decay. This process reduces the energy of the photoelectron slightly
(about 10-40 eV), and the energy difference can either be absorbed by the atom
or added to the energy of the Auger electron. At the end, the atom becomes
doubly ionised. 

(b) Photoemission accompanied by a shake-off event. In the vicinity of a
hole, a vacancy and a free electron are created. The shake electron has
around 10-100 eV energy, and the atom becomes doubly ionised. 

(c) Photoemission accompanied by double Auger decay, which may proceed via
different mechanisms, and may result in the triple ionisation of the
atom. These mechanisms include Auger cascading, single Auger decay
combined with shake-off emission, and virtual inelastic scattering (for a
full description see Amusia {\em et~al.}, 1992). In this case, the energies of
the two ejected electrons are asymmetrically distributed. For instance,
for Ne (transition $1s^{-1}\rightarrow2s^{-2}2p^{-1}+q_1+q_2$) the total available
kinetic energy is about 650 eV, and the most probable case is shaking off a
slow electron and Auger emission of a fast electron with $E_{shake-off} \ll
E_{Auger}$ (Amusia {\em et~al.}, 1992).\\ 

In the initial damage model of proteins (Neutze {\em et~al.}, 2000) only
outer-shell photoionisation and inner-shell photoionisation with a
subsequent Auger emission were taken into account. In the present paper we
focus on photoionisation mechanisms, including shake-up with a subsequent
satellite photoionisation (case a) and shake-off (cases b and c),
which under certain conditions may produce multiple ionisations in the
atom (case c). We investigate in detail what effects these processes may
have on model compounds in order to assess radiation induced damage
processes in biological samples. We calculate the different shake
contributions for carbon, nitrogen, and oxygen atoms in a model peptide
(see Fig.\ 1a). Shake effects for sulphur are estimated in separate
calculations for the three most common chemical environments of sulphur in
proteins (Cys, Met, Cys-S-S-Cys; Figs.\ 1b-d). 

\section*{Results} 

Fig.\ 1 shows the model compounds used in the calculations. In a recent
polymer study (Nakayama {\em et~al.}, 1999), experimentally observed 
differences in
intensity for ester and carbonyl oxygen main lines could be explained
directly from consideration of the main line intensities, where the main
line intensity was estimated from the overlap obtained from the $n = 0$
term in equation (\ref{pn}). This has the clear advantage for inherently
large molecules, such as polymers or polypeptides, that main-line losses
can be estimated without consideration of all the shake states to which
intensity is distributed. For large molecules, the number of shake states
goes beyond that which can readily be calculated at the configuration
interaction level of theory. In a Gly-Gly -Gly tripeptide (Fig.\ 1a), the
central glycine unit can be expected to display a shake spectrum which is
similar to that of a residue within a longer polypeptide chain. To
investigate effects arising from the truncations of the polypeptide chain,
shake effects were calculated for all atoms in the model, including the
atoms from the terminal carboxyl and amino functional groups. \

There are six {\bf carbon} atoms in the peptide model, showing very
similar overlaps between the initial unionized state, and the core ionized
ground state (cf.\ Table 1). This overlap is $\sim0.85$ for all carbons,
except for the carbon at the terminal carboxylic acid group, which has an
overlap of $\sim0.87$. The intensity loss from the main line is therefore
close to ca.\ 30\% for all carbon atoms in these models (cf.\ Table 1). 

The {\bf oxygen} shake contribution is generally the largest. In
particular, the carbonyl oxygens display large shake effects, in agreement
with the general behaviour for organic molecules. Here, strong shake is
found for the two carbonyl oxygens in the peptide links, as well as for
the double-bonded oxygen in the terminal carboxylate functional group. The
overlap for these three oxygens is $\sim0.78$ (cf.\ Table 1),
corresponding to a total shake intensity loss of ca.\ 40\% (cf.\ Table
1). The terminal hydroxyl oxygen has a slightly larger overlap of
$\sim0.82$, corresponding to the main line intensity loss of ca.\ 33\%. 

There are two {\bf nitrogens} in the peptide links, and one in a terminal
amino group. The shake effect for all these atoms is similar: the overlap
is $\sim0.90$ (Table 1) and the main line loss is ca.\ 20\% (cf.\ Table
1). Nitrogen shows the smallest shake contribution, and the results agree
well with data on a polyimide polymer (Nakayama {\em et~al.}, 1999), 
which also showed little nitrogen shake effect.

{\bf Sulphur} shake effects have been calculated for three different
common chemical environments (Fig.\ 1b-d and Table 2). The largest loss
was found for a disulphide bridge (R-S-S-R, 25 \%), and the smallest one
for the thiol terminal group (R-SH, 21 \%).

The initial models used for the sulphur calculations were smaller than the
models used for the other atom types. This could, in principle, be a cause
of concern, as these type of calculations are known to underestimate shake
contributions in extended systems. In order to test the stability of the
numerical results, additional calculations were performed on models with
increasing chain lengths ( $CH_3-SH$, $CH_3-CH_2-SH$, and
$CH_3-CH_2-CH_2-SH$ ). The results indicated negligible drift (not shown). 

\section*{Discussion}

Radiation damage prevents the structural determination of single
biomolecules and other non-repetitive structures (like cells) at high
resolutions in classical electron or X-ray scattering experiments
(Henderson, 1990; Henderson, 1995). Analysis of time-dependent components in
damage formation suggests that the conventional damage barrier can be
substantially extended at extreme dose rates and ultrashort exposure times
(Neutze {\em et~al.}, 2000; Hajdu, 2000; Hajdu {\em et~al.}, 2000;
Hajdu \& Weckert, 2001; Ziaja {\em et~al.}, 2001). A quantitative description of damage formation and a
detailed analysis of the ionisation dynamics of the sample are crucial for
planning experiments at future X-ray lasers. Results described in this
paper give the first assessments of shake contributions to the ionisation
of carbon, nitrogen, oxygen and sulphur atoms within the most common
chemical environments in a protein molecule. 
In previous applications, the calculated shakeup intensities were usually
a few percentage units too high, so we judge the present results to
provide upper bounds to the total shake intensities. 

Data for noble gases show that shake-up and shake-off effects were of
comparable magnitude (Svensson {\em et~al.}, 1988; Armen {\em et~al.}, 1985; 
Wark {\em et~al.}, 1991). Calculations on carbon, oxygen and nitrogen by 
Mohammedein {\em et~al.}, 1993 and El-shemi \& Hassan, 1997 show that photoemission 
accompanied by double Auger decay has nearly zero probability for these light 
elements. Such triple
ionisations can thus be neglected in the ionisation dynamics of C, O, N
compounds. Sulphur, on the other hand, behaves differently, and the
probability of multiple ionisation ($n>2$) after the ejection of a K-shell
photoelectron is larger. The average charge left in sulphur after a single
K-shell photoionisation is estimated to be about 4 (Mohammedein {\em et~al.}, 
1993). For instance, for S: $P(2)\sim5\%$, $P(4)\sim34\%$,
$P(6)\sim3\%$ (Mohammedein {\em et~al.}, 1993; El-shemi \& Hassan, 1997). This 
implies that for sulphur, multiple ionisation mechanisms should be taken into 
consideration in order to obtain 
more accurate predictions for the ionisation dynamics of proteins. 

The energy of the photoelectrons, which emerge from shake-up processes is
similar to the energy of photoelectrons released during outer-shell
ionisation (Nakayama {\em et~al.}, 1999). In these calculations the deviation 
was within 10-40 eV. As a consequence, these electrons can leave a
submicroscopic sample in the same manner as outer-shell
photoelectrons. The inelastic mean free path of these electrons is of the
order of a few hundred \AA ngstroms (Ashley, 1990 and references therein; Ziaja
{\em et~al.}, 2001), 
and thus the damage by the departing electrons will
only have to be considered in large samples where such electrons may
become trapped and deposit energy. 

Low energy shake-off electrons will behave similarly to Auger electrons,
and may cause secondary ionisation in the sample. A detailed analysis of
secondary electron cascades elicited by low energy electrons can be found
in Ziaja {\em et~al.}, 2001. 

\section*{Materials and methods}

The average total shake contributions were estimated from quantum chemical
(QC) calculations, using the semi-empirical INDO/S-CI program ZINDO,
developed by Zerner and co-workers (Ridley \& Zerner, 1976; Bacon \& Zerner,
1979; Zerner {\em et~al.}, 1980),
with standard parameterisation. A single determinant description was used
for the initial (unionized) state of the system, while the excited
core-ionized states were obtained from Configuration-Interaction
calculations with Single excitations (CIS). The equivalent core
approximation was used to represent the core ionized species. 

The calculations are based on the sudden approximation (Aoberg, 1967)
valid for high energy photons, in which the intensity of a given shake
line is proportional to the square of the overlap 
$\langle \Psi_0 | \Phi_n \rangle$~:
\eq 
P_n = | \langle \Psi_0 | \Phi_n \rangle | ^2 , 
\label{pn} 
\eqx 
where $\Psi_0$ is the wave function of the N-1 remaining electrons in the
neutral molecule, and $\Phi_n$ is the (N-1)-electron wave function of the
{\it n}th state of the ionized system. These intensities were calculated
with the SHAKEINT (Lunell, 1987; Lunell \& Keane, 1988) program package. 
This method has been successfully applied to a wide range of systems, 
including organic molecules (c.f. Lunell {\em et~al.}, 1978; Sj\"ogren 
{\em et~al.}, 1992), C60 (Enkvist {\em et~al.}, 1995), polymers (PMDA-ODA 
polyimide) (Nakayama {\em et~al.}, 1999), and adsorbates on metals 
(CO adsorbed on Cu(100) surfaces) (Persson {\em et~al.}, 2000). 
\section*{Acknowledgements} 
We thank Svante Svensson for illuminating discussions and David van der Spoel
for excellent comments. This work was
sponsored by the Swedish Natural Science Research Council (NFR), the
Swedish Research Council for Engineering Sciences (TFR), the EU-BIOTECH
Programme and in part by the Polish Committee for Scientific Research with
grants Nos.\ 2 P03B 04718, 2 P03B 05119 to B. Z. A.\ S.\ was supported by
a STINT distinguished guest professorship. B.\ Z.\ is supported by the
Wenner-Gren Foundations. 


\newpage
\pagestyle{empty} 
\begin{table}[hbpt] 
\noindent 
Table 1.\ {\it Overlaps $\left(\langle \Psi_0 | \Phi_0 \rangle\right)$ 
and average total shake electron intensity 
$\left(1-|\langle \Psi_0 | \Phi_0 \rangle|^2\right)$. 
Values were calculated for all carbon, oxygen, and nitrogen atoms
ocurring in the model polypeptide (Fig.\ 1a).}
\begin{center} 
\begin{tabular}{|c|c|c|} 
\hline \hline 
Atom & Overlap &Average shake intensity\\ 
\hline 
C1 &$0.853$&27\%\\ 
C2 &$0.858$&\\ 
C3 &$0.850$&\\ 
C4 &$0.857$&\\ 
C5 &$0.853$&\\ 
C6 &$0.869$&\\ 
\hline 
O1 &$0.776$&38\%\\ 
O2 &$0.773$&\\ 
O3 &$0.785$&\\ 
O4 &$0.823$&\\ 
\hline 
N1 &$0.902$&19\%\\ 
N2 &$0.896$&\\ 
N3 &$0.897$&\\ 
\hline \hline 
\end{tabular} 
\end{center} 
\end{table} 
\newpage 
\pagestyle{empty} 
\begin{table}[hbpt] 
\noindent 

Table 2.\ {\it Total shake contributions $\left(1-|\langle \Psi_0 |
\Phi_0 \rangle|^2\right)$ 
calculated for sulphur in three different chemical environments.}
\begin{center} 
\begin{tabular}{|c|c|} 
\hline \hline 
Molecule&Shake intensity\\
\hline 
$CH_3CH_2-SH$		& 21\%		\\ 
$CH_3CH_2-S-CH_2CH_3$ 	& 22\%		\\ 
$CH_3CH_2-S-S-CH_2CH_3$ & 25\%		\\ 
\hline
Average & 23\%\\
\hline \hline 
\end{tabular} 
\end{center} 
\end{table} 
\newpage
\thispagestyle{empty} 

Figure 1.\ {\it The model polypeptide for calculating the shake
contributions
for carbon, nitrogen, and oxygen atoms (Fig.\ 1a). Shake effects for
sulphur 
were estimated in separate calculations for the three most common chemical
environments of sulphur in proteins (Cys, Met, Cys-S-S-Cys; Figs.\ 1b-d).}

%
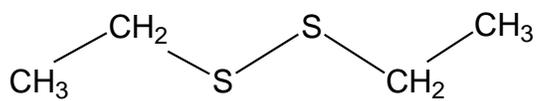
\end{document}